\documentclass[conference]{IEEEtran}
\usepackage{amsmath,amssymb,amsfonts}
\usepackage{algorithmic}
\usepackage{graphicx}
\usepackage{textcomp}
\usepackage{xcolor}
\usepackage{balance}
\usepackage{tabularx}  
\usepackage{booktabs}  
\usepackage{fancyhdr}
\usepackage{url}
\usepackage{hyperref}
\usepackage[square,comma,sort&compress,numbers]{natbib} 
\usepackage{threeparttable,multirow,adjustbox,threeparttable}
\usepackage[symbol]{footmisc}
\usepackage{amsmath, amssymb, amsthm, bm, algorithm, algorithmic}

\hypersetup{
    colorlinks=true,        
    linkcolor=blue,         
    citecolor=blue,         
    filecolor=blue,         
    urlcolor=blue           
}

\usepackage{caption}
\begin{document}
\title{scASDC: Attention Enhanced Structural Deep Clustering for Single-cell RNA-seq Data}
\author{
\IEEEauthorblockN{Wenwen Min$^{1}$, Zhen Wang$^1$, Fangfang Zhu$^2$, Taosheng Xu$^3$, Shunfang Wang$^{1}$}\\
\IEEEauthorblockA{
$^1$School of Information Science and Engineering, Yunnan University, Kunming 650091, Yunnan, China\\ 
$^2$College of Nursing Health Sciences, Yunnan Open University,  Kunming 650599, Yunnan, China\\
$^3$ Hefei Institutes of Physical Science, Chinese Academy of Sciences, Hefei 230031, China
}				
}
\maketitle
\thispagestyle{fancy}
\let\thefootnote\relax\footnotetext{*Corresponding author: minwenwen@ynu.edu.cn}
\begin{abstract}
Single-cell RNA sequencing (scRNA-seq) data analysis is pivotal for understanding cellular heterogeneity. 
However, the high sparsity and complex noise patterns inherent in scRNA-seq data present significant challenges for traditional clustering methods. 
To address these issues, we propose a deep clustering method, Attention-Enhanced Structural Deep Embedding Graph Clustering (scASDC), which integrates multiple advanced modules to improve clustering accuracy and robustness. 
Our approach employs a multi-layer graph convolutional network (GCN) to capture high-order structural relationships between cells, termed as the graph autoencoder module. 
To mitigate the oversmoothing issue in GCNs, we introduce a ZINB-based autoencoder module that extracts content information from the data and learns latent representations of gene expression. These modules are further integrated through an attention fusion mechanism, ensuring effective combination of gene expression and structural information at each layer of the GCN. 
Additionally, a self-supervised learning module is incorporated to enhance the robustness of the learned embeddings. Extensive experiments demonstrate that scASDC outperforms existing state-of-the-art methods, providing a robust and effective solution for single-cell clustering tasks. 
Our method paves the way for more accurate and meaningful analysis of single-cell RNA sequencing data, contributing to better understanding of cellular heterogeneity and biological processes.
All code and public datasets used in this paper are available at \url{https://github.com/wenwenmin/scASDC} and \url{https://zenodo.org/records/12814320}.
\end{abstract}


\section{Introduction}
Cell clustering is one of the most critical tasks in scRNA-seq data analysis \cite{yu2023topological,grabski2023significance}. 
However, due to the limitations of sequencing technology, scRNA-seq data often exhibits high sparsity and complex noise patterns \cite{villani2017single,zheng2023tsimpute}. 
Traditional clustering methods \cite{min2023structured,min2019group,wang2024graph} such as k-means and hierarchical clustering, which rely on similarity measures, struggle to meet the demanding requirements of single-cell data clustering due to these inherent challenges.

To better capture the unique characteristics of scRNA-seq data, deep learning-based clustering algorithms have emerged and been widely applied \cite{molho2024deep}. 
Examples of these algorithms include DESC \cite{desc}, scDCC \cite{tian2021model}, and scDeepCluster \cite{tian2019clustering}. 
These methods typically utilize autoencoders to learn low-dimensional representations of the data, which helps in clustering and distribution analysis. 
However, these approaches mainly focus on gene expression information and often overlook the relationships between cells, i.e., the structural information within the data. 
Given the sparsity and noise in single-cell data, relying solely on gene expression information for clustering can lead to suboptimal results. 
It is crucial to incorporate both gene expression and structural information for more robust cell clustering.

To address these limitations, recent studies have explored the integration of graph-based methods to capture cell-cell relationships and enhance clustering performance \cite{fang2024contrastive,min2024d,huo2021caegcn}. 
Techniques such as scGNN \cite{wang2021scgnn}, scGAE \cite{luo2021topology}, scTAG \cite{yu2022zinb} and scDSC \cite{scDSC} leverage graph convolutional networks (GCNs) to extract structural information, thereby improving clustering accuracy. 
However, these methods often suffer from oversmoothing when dealing with large-scale data, leading to the loss of critical features embedded in the gene expression matrix \cite{kipf2016semi}.

This paper proposes an innovative approach named Attention-Enhanced Structural Deep Embedding Graph Clustering (scASDC). 
Our method employs a multi-layer GCN to capture high-order structural relationships in single-cell data, termed as the graph autoencoder module. 
To mitigate the oversmoothing issue in GCNs, we introduce a ZINB-based autoencoder module to extract content information from the data, learning the latent representations of gene expression data \cite{miao2018desingle}. 
By integrating these two modules, our approach combines gene expression and structural information through an attention fusion mechanism, enhancing clustering performance significantly.

The core components of our method include four modules: the ZINB-based autoencoder module, the graph autoencoder module, the attention fusion module, and the self-supervised module. 
After preprocessing, the single-cell RNA sequencing data matrix is input into the ZINB-based autoencoder module. 
This module learns a mapping from the data space to a low-dimensional feature space, capturing the content information of the data and producing a low-dimensional latent representation. 
The ZINB decoder can then reconstruct the gene expression matrix and the data distribution.

Simultaneously, we construct a cell-cell graph based on the single-cell RNA sequencing matrix and use it as the initial input for the graph autoencoder module. 
This module, composed of multiple layers of GCNs, effectively extracts high-order structural information between cells. 
A key innovation of our approach is the use of an attention fusion mechanism that iteratively integrates the output of the ZINB-based autoencoder and the graph autoencoder module at each layer of the GCN. 
This layer-by-layer embedding process ensures that both feature representations and structural information are combined into a unified representation.

Finally, a self-supervised strategy integrates the ZINB-based autoencoder module and the graph autoencoder module into a unified framework. 
This allows for end-to-end clustering training and synchronous optimization of both modules, ensuring that the learned representations are robust and effective for downstream tasks such as single-cell clustering.

The main contributions of our proposed method are:
\begin{itemize}
    \item We propose a single-cell deep graph clustering method called scASDC, which enhances cell clustering by combining GCN-based structural information with ZINB-based gene expression embeddings.
    \item Our method integrates an attention fusion module that selectively emphasizes important features, thereby improving the integration of gene expression and structural data. Additionally, a self-supervised learning module is adopted to enhance the robustness and accuracy of the learned cell embeddings.
    \item We compare our method with seven competitive baseline methods on six scRNA-seq datasets. The results demonstrate that scASDC outperforms the existing state-of-the-art techniques.
\end{itemize}

\section{MATERIALS AND METHODS}\label{MATERIALS AND MRTHODS}
\subsection{Dataset and pre-processing}\label{Dataset and pre-processing}
We collect six scRNA-seq datasets across multiple platforms, encompassing various organs and cell types, including kidney, limb, and diaphragm cells from both humans and mice. 
These datasets include `QS Limb Muscle', `Adam', `QS Diaphragm', `QS Trachea', `Romanov', and `QX Limb Muscle'. 
Detailed information on these datasets is provided in \autoref{tab-1}.
\begin{figure*}[htp]%
\centering
\centering
\includegraphics[width=0.85\textwidth]{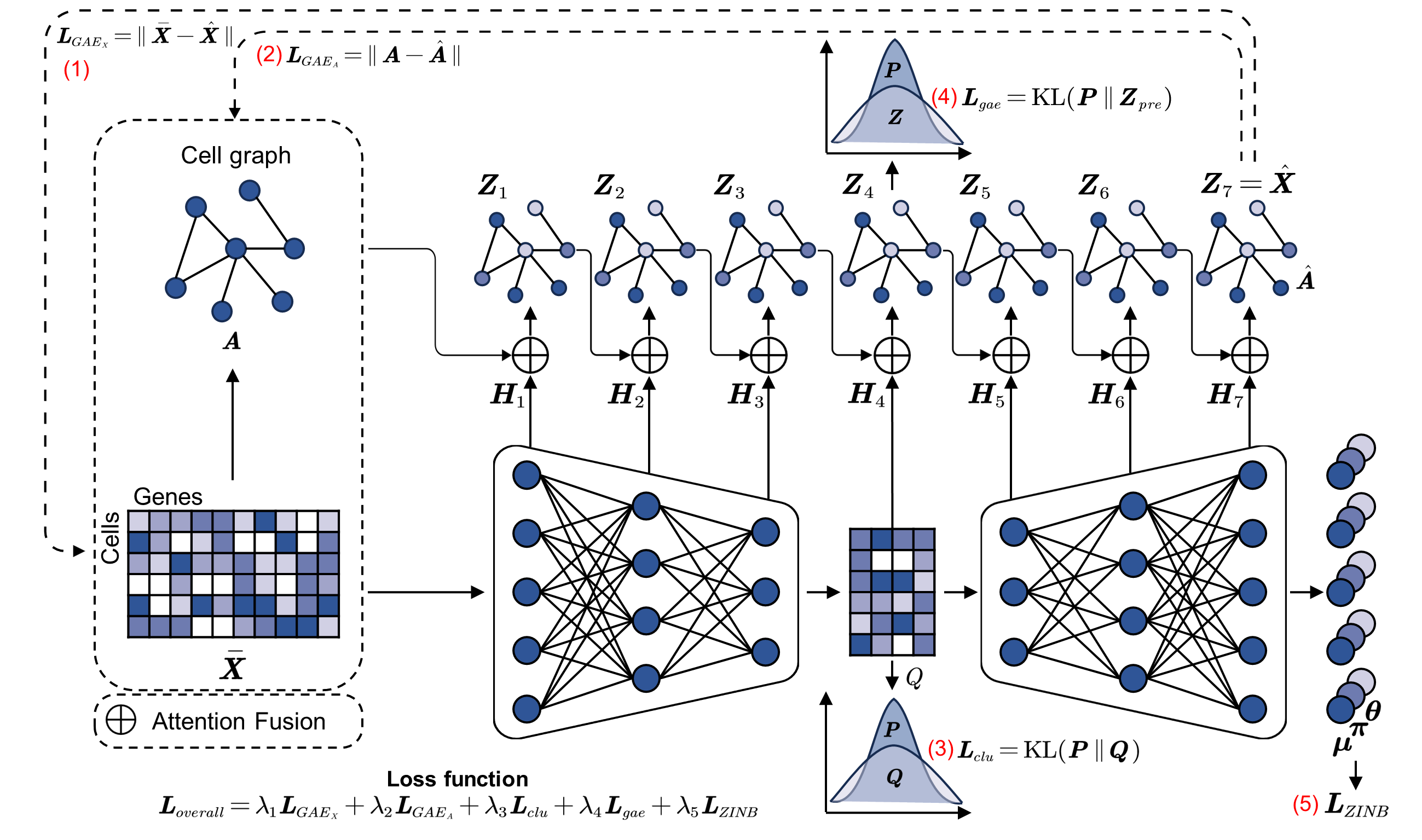}
\caption{Model framework of scASDC.
The scASDC framework is primarily composed of a ZINB-based autoencoder module, a graph autoencoder module, an attention fusion module, and a self-supervised module. The framework leverages the attention mechanism to embed and transmit the outputs of the two autoencoder modules layer by layer. This process ensures that the obtained embedding representation retains both the content and structural information of the original data. Additionally, the self-supervised module integrates multiple networks into a single framework, facilitating end-to-end synchronous updates.
In this framework, $\mathbf{\bar{X}}$ represents the input data, and $\mathbf{A}$ denotes the original cell graph. $\mathbf{H}_l$ and $\mathbf{Z}_l$ indicate the outputs of the ZINB-based autoencoder module and the $l$-th layer of the graph autoencoder module, respectively. The parameters $\pi$, $\mu$, and $\theta$ correspond to the three parameters of the ZINB distribution.}
\label{fig1}
\end{figure*}
The original gene expression matrix obtained from the scRNA-seq data was preprocessed as follows:
We use the scRNA-seq gene expression data matrix $X \in \mathbb{R}^{n \times g}$ as input, where $n$ represents the number of cells and $g$ represents the number of genes.
First, we filter out genes that are not expressed in any cells to reduce the impact of sparsity inherent in scRNA-seq data.
Next, we use the Scanpy package \cite{wolf2018scanpy} to normalize the data, converting discrete data into continuous data.
Finally, the genes are ranked according to their normalized variance values, and the top $d$ highly variable genes are selected to obtain the normalized data matrix $\bar{X} \in \mathbb{R}^{n \times d}$.

To capture the structural relationships between cells, we use the preprocessed gene expression matrix $\bar{X}$ to construct a K-nearest neighbor (KNN) graph $A \in \mathbb{R}^{n \times n}$ \cite{baran2019metacell}. 
In the KNN graph, each node represents a cell, and the edges between nodes represent the relationships between cells.

\begin{table}[htp]
\caption{Summary of scRNA-seq datasets used in this study.}\label{tab-1}
\begin{adjustbox}{width=0.5\textwidth}
\begin{tabular}{l|c|c|c|c|c|c}
\hline
Datasets        & \#Cell  & \#Gene  & \#Class & Organ        & Platform   &Ref.                  \\ \hline
QS Limb Muscle & 1090  & 23341 & 6     & Limb Muscle  & Smart-seq2 &\cite{schaum2018single}      \\
Adam           & 3660  & 23797 & 8     & Kidney       & Drop-seq   &\cite{adam2017psychrophilic} \\
QS Diaphragm   & 870   & 23341 & 5     & Diaphragm    & Smart-seq2 &\cite{schaum2018single}      \\
QS Trachea     & 1350  & 23341 & 4     & Trachea      & Smart-seq2 &\cite{schaum2018single}      \\
Romanov        & 2881  & 21143 & 7     & Hypothalamus & SMARTer    &\cite{romanov2017molecular}  \\ 
Qx Limb Muscle & 3909  & 23341 & 6     & Limb Muscle  & 10x        &\cite{schaum2018single}      \\ \hline
\end{tabular}
\end{adjustbox}
\end{table}

\subsection{The proposed scASDC method}\label{scASDC framework}

\subsubsection{ZINB-based autoencoder}\label{ZINB-based autoencoder}
The autoencoder module consists of symmetrical encoder and decoder components. Specifically, assuming the encoder module of the autoencoder has $L$ layers, the input to the $l$-th layer is $\mathbf{H}_{l-1}$, and its output can be represented as:
\begin{equation}
    \mathbf{H}_l=\phi(\mathbf{W}_l \mathbf{H}_{l-1}+\mathbf{b}_l)
\end{equation}
where the activation function $\phi$ can be flexibly chosen according to the specific application. We select the ReLU function as the activation function. $\mathbf{W}_l$ represents the weights of the $l$-th layer, and $\mathbf{b}_l$ represents the bias vector of the $l$-th layer. It is noteworthy that the input to the first layer is the preprocessed gene expression matrix $\bar{X}$, and the output of the final layer is the reconstructed gene expression matrix, whose dimensions are the same as the original input matrix.

To capture the global probabilistic structure of the data, we integrate the ZINB model into the decoder structure of the autoencoder. Specifically, we connect three independent fully connected layers to the final layer of the autoencoder to estimate the three parameters of the ZINB distribution: the drop out parameter $\pi$, the dispersion parameter $\theta$, and the mean parameter $\mu$:
\begin{equation}
    \Pi=\operatorname{sigmoid}\left(W_{\pi} \mathbf{H}_L\right)
\end{equation}
\begin{equation}
    M=\operatorname{diag}\left(S_i\right)\times \exp \left(W_{\mu} \mathbf{H}_L\right)
\end{equation}
\begin{equation}
    \Theta= \exp \left(W_{\theta} \mathbf{H}_L\right)
\end{equation}
where the size factor $S_i$ is calculated in the data preprocessing stage by dividing the total expression of each cell by the total expression of a reference cell, $W_{\pi}$, $W_{\mu}$, $W_{\theta}$ represent the weight parameters. The ZINB distribution uses the above three parameters to model the original scRNA-seq data and reconstruct its data distribution:
\begin{equation}
    NB(\bar{X} \mid \mu, \theta)=\frac{\Gamma(\bar{X}+\theta)}{\bar{X}!\Gamma(\theta)}\left(\frac{\theta}{\theta+\mu}\right)^{\theta}\left(\frac{\mu}{\theta+\mu}\right)^{\bar{X}}
\end{equation}
\begin{equation}
    ZINB(\bar{X} \mid \pi, \mu, \theta)=\pi \delta_{0}(\bar{X})+(1-\pi) NB(\bar{X})
\end{equation}

To guide the model in learning more effective representations, we hope that the reconstructed ZINB distribution is as close as possible to the original data distribution. Therefore, we define the final loss function of the ZINB-based autoencoder module as the negative log-likelihood of the ZINB distribution:
\begin{equation}
    \mathcal{L}_{ZINB}=-\log (ZINB(\bar{X} \mid \pi, \mu, \theta))
\end{equation}

\subsubsection{Graph Autoencoder Module}\label{Graph Autoencoder Module}
To capture the structural information between cells, we introduce a graph autoencoder module that uses a graph convolutional network (GCN) as its backbone, taking the KNN graph $\mathbf{A}$ as input to effectively extract structural information while fusing content and structural information in a heterogeneous structure.

\begin{table*}[t]
\caption{Performance of Our Method Compared to Baseline Methods Across Six scRNA-seq Datasets (Best values in bold).}\label{tab-2}
\begin{adjustbox}{width=1\textwidth}
\fontsize{6}{6}\selectfont
\begin{tabular}{l|l|c|c|c|c|c|c|c|c}
\hline
                     \textbf{Metrics} & \textbf{Datasets}         & \textbf{scASDC~(ours)}          & \textbf{DESC}   & \textbf{SDCN}   & \textbf{scDeepCluster} & \textbf{DCA}    & \textbf{scDSC}  & \textbf{DEC}    & \textbf{AttentionAE-sc}  \\ \hline
\multirow{7}{*}{NMI} 
                     & QS\_limb\_muscle & \textbf{0.9413±0.004} & 0.8391±0.019 & 0.9333±0.046 & 0.7489±0.009  & 0.6217±0.026 & 0.9331±0.017 & 0.9297±0.003 & 0.9026±0.051          \\
                     & Adam             & \textbf{0.8283±0.017} & 0.7991±0.010 & 0.7786±0.017 & 0.8116±0.001  & 0.5713±0.006 & 0.8063±0.018 & 0.7548±0.001 & 0.7280±0.022          \\
                     & QS\_Diaphragm    & \textbf{0.9467±0.008} & 0.8961±0.060 & 0.9435±0.021 & 0.8155±0.001  & 0.8124±0.010 & 0.9375±0.010 & 0.8595±0.003 & 0.9224±0.013          \\
                     & QS\_Trachea      & \textbf{0.8134±0.004} & 0.5726±0.016 & 0.7328±0.033 & 0.6015±0.036  & 0.6759±0.005 & 0.7058±0.020 & 0.7380±0.000 & 0.7266±0.041          \\
                     & Romanov          & \textbf{0.7208±0.011} & 0.6149±0.011 & 0.6976±0.002 & 0.6189±0.002  & 0.5145±0.007 & 0.6969±0.014 & 0.6132±0.001 & 0.7137±0.054          \\
                     & QX\_Limb\_Muscle & \textbf{0.9665±0.004} & 0.7970±0.009 & 0.9329±0.001 & 0.9550±0.004  & 0.7189±0.029 & 0.9556±0.022 & 0.8487±0.048 & 0.9476±0.016          \\ \hline
\multirow{7}{*}{ARI} 
                     & QS\_limb\_muscle & \textbf{0.9656±0.002} & 0.6499±0.029 & 0.9572±0.035 & 0.6801±0.001  & 0.6507±0.041 & 0.9635±0.007 & 0.9619±0.002 & 0.9428±0.018          \\
                     & Adam             & \textbf{0.7651±0.022} & 0.6750±0.020 & 0.6861±0.029 & 0.7270±0.001  & 0.4147±0.008 & 0.7185±0.030 & 0.6571±0.002 & 0.6729±0.248          \\
                     & QS\_Diaphragm    & \textbf{0.9726±0.005} & 0.8564±0.158 & 0.9668±0.019 & 0.6996±0.001  & 0.8491±0.012 & 0.9654±0.004 & 0.9293±0.002 & 0.9592±0.012          \\
                     & QS\_Trachea      & \textbf{0.8662±0.003} & 0.2692±0.023 & 0.7354±0.062 & 0.5287±0.134  & 0.6463±0.008 & 0.7463±0.068 & 0.8054±0.000 & 0.6882±0.108          \\
                     & Romanov          & \textbf{0.7858±0.007} & 0.4219±0.026 & 0.7380±0.002 & 0.6282±0.001  & 0.4073±0.003 & 0.7344±0.011 & 0.4590±0.001 & 0.7230±0.149          \\
                     & QX\_Limb\_Muscle & \textbf{0.9829±0.003} & 0.5836±0.030 & 0.9491±0.001 & 0.9611±0.001  & 0.5990±0.046 & 0.9702±0.026 & 0.8961±0.109 & 0.9649±0.014          \\ \hline
\end{tabular}
\end{adjustbox}
\end{table*}

\begin{table}[t]
\centering
\caption{Ablation study results measured by NMI and ARI values.} \label{tab-3}
\begin{adjustbox}{width=0.48\textwidth} 
\fontsize{6}{6}\selectfont         
\setlength{\arrayrulewidth}{0.05mm}    
\begin{tabular}{c|l|cccc}
\hline
Metric               & Datasets         & scASDC           & \begin{tabular}[c]{@{}c@{}}scASDC w/o\\      ZINB Loss\end{tabular}  & \begin{tabular}[c]{@{}c@{}}scASDC w/o\\       Attention\end{tabular}  & \begin{tabular}[c]{@{}c@{}}scASDC w/o\\       Graph Loss\end{tabular} \\ \hline
\multirow{7}{*}{NMI} & QS\_Limb\_Muscle & 0.9562±0.008          & 0.9417±0.004                                                              & 0.9427±0.005                                                               & 0.9531±0.014                                                                \\
                     & Adam             & 0.7722±0.004          & 0.7655±0.037                                                              & 0.7497±0.017                                                               & 0.7661±0.037                                                                \\
                     & QS\_Diaphragm    & 0.9561±0.008          & 0.7416±0.008                                                              & 0.9449±0.003                                                               & 0.9367±0.007                                                                \\
                     & QS\_Trachea      & 0.8104±0.032          & 0.8188±0.016                                                              & 0.7926±0.004                                                               & 0.7865±0.035                                                                \\
                     & Romanov          & 0.7022±0.011          & 0.6963±0.010                                                              & 0.7010±0.006                                                               & 0.5492±0.070                                                                \\
                     & QX\_Limb\_Muscle & 0.8730±0.024          & 0.8671±0.021                                                              & 0.8749±0.029                                                               & 0.8703±0.004                                                                \\
                     & Mean             & \textbf{0.8450±0.103}  & \textbf{0.8052±0.089}                                                      & \textbf{0.8343±0.102}                                                       & \textbf{0.8103±0.148}                                                       \\ \hline
\multirow{7}{*}{ARI} & QS\_limb\_muscle & 0.9765±0.005          & 0.9706±0.002                                                              & 0.9709±0.003                                                               & 0.9719±0.006                                                                \\
                     & Adam             & 0.6931±0.003          & 0.6885±0.093                                                              & 0.6718±0.220                                                               & 0.6827±0.074                                                                \\
                     & QS\_Diaphragm    & 0.9779±0.005          & 0.6866±0.006                                                              & 0.9706±0.002                                                               & 0.9649±0.006                                                                \\
                     & QS\_Trachea      & 0.7689±0.045          & 0.8104±0.016                                                              & 0.7611±0.003                                                               & 0.7593±0.040                                                               \\
                     & Romanov          & 0.7651±0.007          & 0.7618±0.006                                                              & 0.7660±0.005                                                               & 0.4131±0.038                                                                \\
                     & QX\_Limb\_Muscle & 0.7872±0.021          & 0.7867±0.069                                                              & 0.7888±0.036                                                               & 0.7734±0.003                                                                \\
                     & Mean             & \textbf{0.8281±0.119}  & \textbf{0.7841±0.104}                                                      & \textbf{0.8215±0.122}                                                       & \textbf{0.7609±0.206}                                                       \\ \hline
\end{tabular}
\end{adjustbox}
\end{table}

Specifically, assuming that the graph autoencoder module has $L$ layers, and the input of the $l-$th layer is $\boldsymbol{Z}_{l-1}$, its output can be expressed as:
\begin{equation}
    \mathbf{Z}_l=\phi \left( \hat{\mathbf{D}}^{-\frac{1}{2}}\left( \mathbf{A}+\mathbf{I} \right) \hat{\mathbf{D}}^{-\frac{1}{2}}\mathbf{Z}_{l-1}\mathbf{U}_{l-1} \right) 
\end{equation}
where $\mathbf{I}$ is the unit diagonal matrix, $\hat{\mathbf{D}}$ is the degree matrix, $\hat{\mathbf{D}}_{ii}=\sum_j{\left( \mathbf{A}_{ij}+\mathbf{I}_{ij} \right)}$, $\mathbf{U}_{l-1}$ represents the weight of the $l-1_{th}$ layer, and the normalized adjacency matrix $\hat{\mathbf{D}}^{-\frac{1}{2}}\left( \mathbf{A}+\mathbf{I} \right) \hat{\mathbf{D}}^{-\frac{1}{2}}$ propagates $\mathbf{Z}_{l-1}$ to obtain the new representation $\mathbf{Z}_l$.
To obtain a richer representation, we embed the output of the ZINB-based autoencoder and the output of the graph autoencoder through an attention fusion module, performing layer-by-layer heterogeneous fusion embedding. 
This process results in a more robust representation $\mathbf{R}_{l-1}$ that integrates both structural and content information:
\begin{equation}\label{eq-9}
    \mathbf{R}_{l-1}=F_{att}\left( \alpha \mathbf{H}_{l-1}+(1-\alpha )\mathbf{Z}_{l-1} \right) 
\end{equation}
where $\alpha$ is an adjustable weight parameter. The specific implementation of the attention fusion operation $F_{att}$ will be introduced in detail in the attention fusion module below.

To enable the network to learn richer representations and reduce noise interference, we use the integrated representation $\mathbf{R}_{l-1}$ as the input for the GCN to learn high-order discriminative information and generate new representations:
\begin{equation}
    \mathbf{Z}_l=\phi \left( \hat{\mathbf{D}}^{-\frac{1}{2}}\left( \mathbf{A}+\mathbf{I} \right) \hat{\mathbf{D}}^{-\frac{1}{2}}\mathbf{R}_{l-1}\mathbf{U}_{l-1} \right) 
\end{equation}

It is worth noting that the input $\mathbf{R}_0$ of the first layer of the graph autoencoder module is the preprocessed gene expression matrix $\boldsymbol{\bar{X}}$.
In the representation learned by the GCN, the middle layer $\mathbf{Z}_{\frac{L}{2}}$ is selected for subsequent clustering. 
A softmax operation is then performed on this middle layer to predict the probability distribution of the gene expression data:
\begin{equation}
    \mathbf{Z}_{pre}=\mathrm{softmax} \left( \hat{\mathbf{D}}^{-\frac{1}{2}}\left( \mathbf{A}+\mathbf{I} \right) \hat{\mathbf{D}}^{-\frac{1}{2}}\mathbf{R}_{\frac{L}{2}-1}\mathbf{U}_{\frac{L}{2}-1} \right) 
\end{equation}

The representation $z_{ij}\in \mathbf{Z}_{pre}$ can be regarded as the probability that the $i$th sample belongs to the $j$th cluster center, which is used for end-to-end model learning and training of the subsequent self-supervised module. 
To guide the model training direction and make the learned representation more credible, we use the following two reconstruction errors as the loss function of the graph autoencoder module:
\begin{equation}\label{eq-12}
    \mathcal{L} _{GAE_A}=\parallel \mathbf{A}-\hat{\mathbf{A}}\parallel _{F}^{2}
\end{equation}
\begin{equation}\label{eq-13}
    \mathcal{L} _{GAE_X}=\parallel \mathbf{\bar{X}}-\mathbf{Z}_L\parallel _{F}^{2}
\end{equation}
where $\hat{\mathbf{A}}=\mathrm{sigmoid}\left( \mathbf{Z}_{L}^{T}\mathbf{Z}_L \right)$. The output $\mathbf{Z}_L$ from the last layer of the graph autoencoder module is converted into the reconstructed adjacency matrix $\hat{\mathbf{A}}$ through an inner product operation. By minimizing Eq.(\ref{eq-12}), the graph autoencoder module can capture more effective structural information, which improves subsequent clustering performance. The minimization of Eq.(\ref{eq-13}) ensures that the final output $\mathbf{Z}_L$ of the graph autoencoder module retains some expression information from the original single-cell RNA sequencing data, thereby preserving specific expression patterns in the final data representation.

\begin{figure}[h]%
\centering
\includegraphics[width=0.49\textwidth]{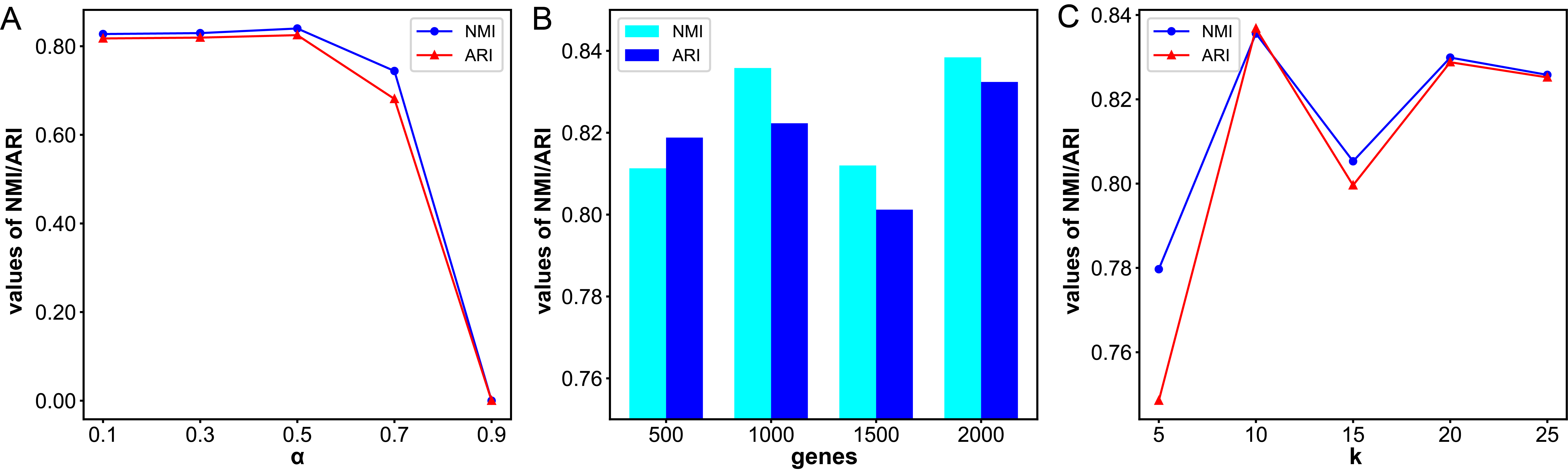}
\caption{The analysis of (A) the average NMI and ARI values with different fusion parameter $\alpha$. (B) Comparison of the average NMI and ARI values with different numbers of genes. (C) the average NMI and ARI values with different neighbor parameter $k$.}
\label{fig-2}
\end{figure}

\subsubsection{Attention Fusion Module}\label{Attention Fusion Module}
The attention mechanism maps a query and a set of key-value pairs to an output, where the query, key, value, and output are all vectors. Specifically, we calculate the dot product of the query and the key, apply the softmax function to obtain the relevant weight of the value, and multiply the value by this weight to get the final result. To simplify the calculation, a set of key-value pairs is computed simultaneously in practice, converting vector calculations into matrix operations. The specific calculation method of the attention mechanism is as follows:
\begin{equation}
    \text { Attention }(\mathbf{Q}, \mathbf{K}, \mathbf{V})=\operatorname{softmax}\left({\mathbf{Q} \mathbf{K}^{T}}\right) \mathbf{V}
\end{equation}
where $\mathbf{Q}, \mathbf{K}, \mathbf{V}$ are the query matrix, key matrix, and value matrix, respectively. Building on this, we introduce a multi-head attention mechanism, which can simultaneously capture information from each subspace at different positions of the data. This ensures that the attention fusion representation contains richer and more robust information. Specifically, this approach performs different attention operations on $\mathbf{Q}, \mathbf{K}, \mathbf{V}$ $M$ times, then executes the attention function in parallel, concatenates the results, and projects them again to obtain the final output. The $i$-th attention operation is expressed as follows:
\begin{equation}
    \text{head}_{i}=\operatorname{Attention}\left(\mathbf{W}_{i}^{Q}\mathbf{Q}, \mathbf{W}_{i}^{K}\mathbf{K}, \mathbf{W}_{i}^{V}\mathbf{V}\right)
\end{equation}
where $\mathbf{W}_{i}^{Q}, \mathbf{W}_{i}^{K}, \mathbf{W}_{i}^{V}$ are transformation matrices responsible for projecting the corresponding query, key, and value matrices, respectively. Assume that the attention module has a total of $L$ layers, and the output of its $l$-th layer is:
\begin{equation}
\begin{split}
    \mathbf{R}_{l}&=F_{\text {att }}(\mathbf{Q}, \mathbf{K}, \mathbf{V})\\
    &=F_{\text {att }}\left(\mathbf{W}^{q} \mathbf{Y}_{l-1}, \mathbf{W}^{k} \mathbf{Y}_{l-1}, \mathbf{W}^{v} \mathbf{Y}_{l-1}\right)\\
    &=\mathbf{W} \cdot \operatorname{Concat}\left(\text{head}_{1}, \ldots, \text{head}_{M}\right)
\end{split}
\end{equation}
Here, $\mathbf{Y}_{l-1} = \alpha \mathbf{H}_{l-1} + (1 - \alpha) \mathbf{Z}_{l-1}$, $\mathbf{W}^{q}, \mathbf{W}^{k}, \mathbf{W}^{v}$ are three transformation matrices, $\mathbf{W}$ is a weight matrix, and $\operatorname{Concat}(\cdot)$ represents the matrix concatenation operation.

\subsubsection{Self-supervised module}\label{Self-supervised module}
We first perform k-means clustering on the intermediate layer output $\mathbf{H}_{\frac{L}{2}}$ of the ZINB-based autoencoder module to obtain a set of initial cluster centers $\{\mu_i\}_{i=1}^{k}$, where $k$ is the predefined number of clusters. Next, we calculate the soft assignment between the embedded representation $\mathbf{H}_{\frac{L}{2}}$ and the cluster centers $\{\mu_i\}_{i=1}^{k}$ using Student's t-distribution to measure similarity. For the $i$-th sample and the $j$-th cluster center, the soft assignment $q_{ij} \in Q$ is calculated as follows:
\begin{equation}
    q_{ij}=\frac{\left( 1+\left\| h_i-\mu _j \right\| ^2/\lambda \right) ^{-\frac{\lambda +1}{2}}}{\sum_{j^{\prime}}{\left( 1+\left\| h_i-\mu _{j^{\prime}} \right\| ^2/\lambda \right) ^{-\frac{\lambda +1}{2}}}}
\end{equation}
where $h_i$ is the $i$-th sample of the embedded representation $\mathbf{H}_{\frac{L}{2}}$, $\lambda $ represents the degree of freedom of Student's t distribution.

Based on the soft cluster distribution $Q$, an auxiliary target distribution is required to supervise the learning of the soft cluster distribution and improve clustering by learning the high-confidence distribution of the target distribution. We use the soft cluster frequencies to calculate a target distribution $p_{ij} \in P$ with higher confidence. The calculation method is as follows:
\begin{equation}
    p_{ij}=\frac{q_{ij}^{2}/g_j}{\sum_{j^{\prime}}{q_{ij^{\prime}}^{2}}/g_{j^{\prime}}}
\end{equation}
where \(g_j = \sum_i q_{ij}\) represents the soft clustering frequency. 
To align the distributions \(Q\) and \(P\), we use the KL divergence loss between the two distributions as the optimization target for this part, aiming to achieve higher quality clustering:
\begin{equation}\label{eq-19}
    \mathcal{L} _{clu}=\mathrm{KL(}P\parallel Q)=\sum_i{\sum_j{p_{ij}}}\log \frac{p_{ij}}{q_{ij}}
\end{equation}
\begin{figure*}[!htp]%
    \centering
    \includegraphics[width=1\textwidth]{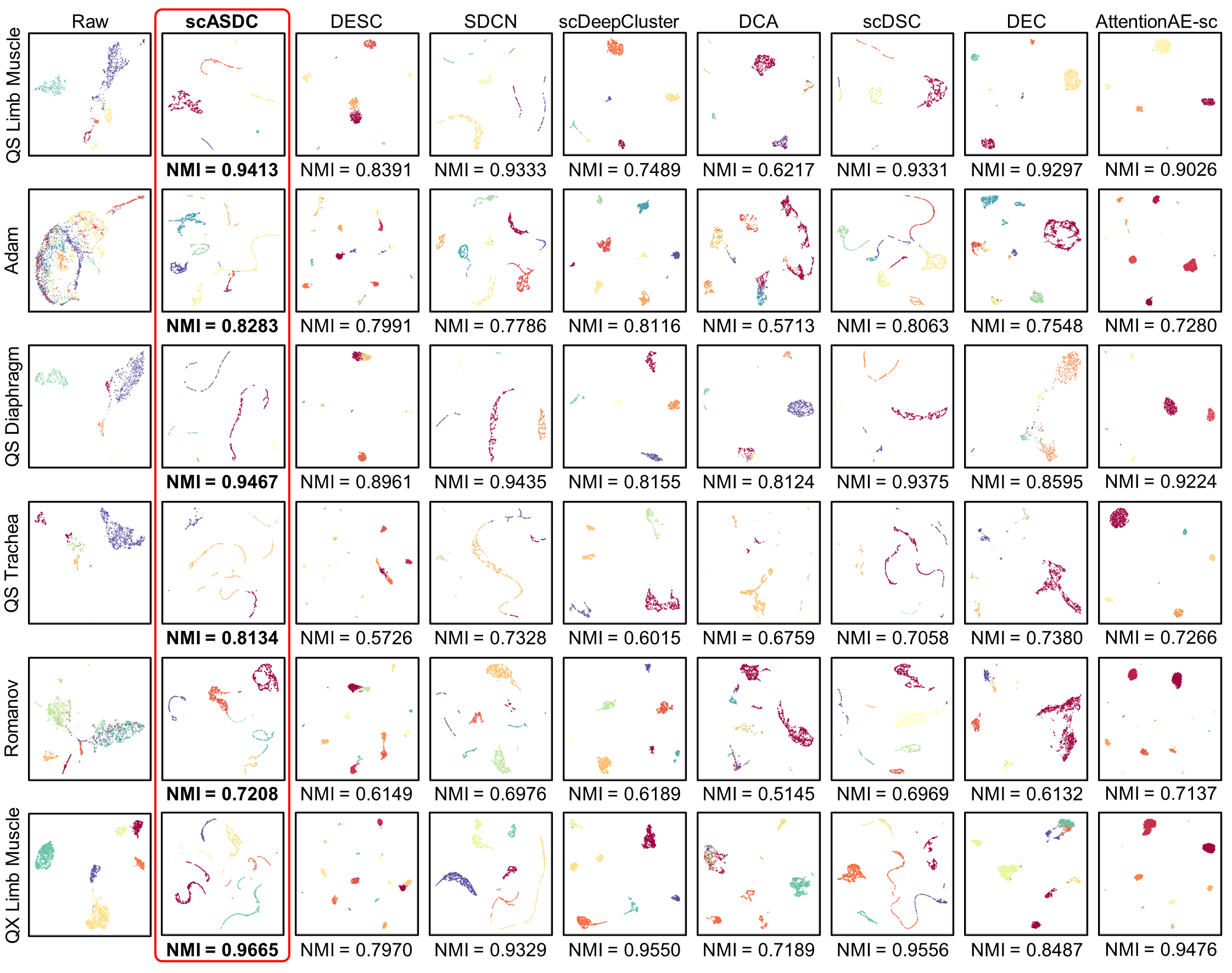}
    \caption{Comparison of UMAP clustering results on six datasets with 2D visualization.}\label{fig-3}
\end{figure*}
Similarly, we can adopt the same self-supervised strategy and use the target distribution $P$ to supervise the learning of the graph autoencoder module:
\begin{equation}\label{eq-20}
    \mathcal{L} _{gae}=\mathrm{KL(}P\parallel Z_{pre})=\sum_i{\sum_j{p_{ij}}}\log \frac{p_{ij}}{z_{ij}}
\end{equation}
where $z_{ij} \in \mathbf{Z}_{pre}$. 
By optimizing the two loss functions from Eq.(\ref{eq-19}) and Eq.(\ref{eq-20}), we integrate the optimization objective into a distribution $P$, making the learned representation more suitable for clustering tasks.

The total loss function of the scASDC model is as follows:
\begin{equation}
    \mathcal{L} =\lambda _1\mathcal{L} _{GAE_X}+\lambda _2\mathcal{L} _{GAE_A}+\lambda _3\mathcal{L} _{clu}+\lambda _4\mathcal{L} _{gae}+\lambda _5\mathcal{L} _{ZINB}
\end{equation}
where $\lambda _1,\lambda _2,\lambda _3,\lambda _4,\lambda _5$ are hyperparameters that measure the importance of each loss function.

\subsection{Evaluation metrics}\label{Evaluation metrics}
We adopt two widely used evaluation metrics Normalized Mutual Information (NMI) and Adjusted Rand Index (ARI) to evaluate the clustering performance of the methods. The larger the values of these metrics, the better the clustering performance of the method.

\section{Experimental results}\label{Experimental results}
\subsection{Implementation details}
For all baselines, default parameters from the original papers
were used, and all experiments were conducted on an NVIDIA
GeForce RTX 3090.
In our scASDC, the weight coefficients of the Loss function are set as follows: $\lambda_1 = 0.5$, $\lambda_2 = 0.01$, $\lambda_3 = 0.1$, $\lambda_4 = 0.01$, and $\lambda_5 = 0.5$.
We pre-train the ZINB-based autoencoder module before the formal training phase. During pre-training, we set the initial learning rate ($\text{lr}$) to 0.001, ran for 100 epochs, and used the Adam optimizer to adjust the learning rate. In the formal training phase, the initial learning rate ($\text{lr}$) was also set to 0.001, with the Adam optimizer used again, $\text{epoch}=200$.
The ZINB-based autoencoder module features a symmetrical structure for its encoding and decoding layers. Each layer of the encoder is configured with 1000, 1000, 4000, and 10 nodes, respectively. In the attention fusion module, we set the number of attention heads to 8.

\subsection{Comparison with baseline methods}
We select the following seven single-cell clustering baseline methods for comparison:
\begin{itemize}
    \item DESC \cite{desc}: It is an unsupervised deep embedding algorithm that clusters scRNA-seq data by iteratively optimizing the clustering target, which can effectively eliminate the batch effect.
    \item SDCN \cite{sdcn}: This method is a structural deep clustering network that utilizes graph convolutional networks to integrate structural information into deep clustering models.
    \item scDeepCluster \cite{tian2019clustering}: scDeepCluster is a deep embedding denoising network that adds a ZINB model to the autoencoder that can simulate the distribution of scRNA-seq data, and then learns feature representation of the data to guide single-cell clustering.
    \item DCA \cite{dca}: DCA is a deep count autoencoder denoising network that takes into account the sparsity of scRNA-seq data as well as dropout events and uses a zero-inflated negative binomial noise model to restore its data distribution.
    \item scDSC \cite{scDSC}: scDSC integrates structural information into single-cell deep clustering, adds a ZINB model to the basic autoencoder, and introduces a GNN module to capture the structural information between cells.
    \item DEC \cite{dec}: It is a deep embedding clustering method that guides the learning of data representation by designing a clustering objective.
    \item AttentionAE-sc \cite{li2023attention}: AttentionAE-sc introduces an attention mechanism to fuse structural embedding and denoising embedding together to obtain a more robust representation, thereby improving clustering performance.
\end{itemize}

\autoref{tab-2} presents the clustering performance results of our scASDC method and seven baseline methods across six scRNA-seq datasets. 
We repeat the experiments five times using 500, 1000, 1500, and 2000 highly variable genes, and calculated the average values. The best highly variable gene setting is selected as the final result. 
Bold values in the table indicate the average indices with the best clustering performance. 
As shown in the table, our method consistently outperforms the baseline methods, achieving the best average indices across all six datasets. 
Notably, the clustering indices on the QX Limb Muscle dataset reach 0.9665 and 0.9829, respectively. 
Additionally, we observed that deep graph embedding clustering methods such as SDCN and scDSC also demonstrat strong clustering performance, indicating that embedding the graph structure effectively captures important structural information in scRNA-seq data, positively impacting clustering results.

\subsection{Parameter analysis}
\subsubsection{Influence of Balance Parameter $\alpha$}
We conduct experiments on six datasets and averaged the results. \autoref{fig-2}A shows the impact of the balance parameter $\alpha$ in Eq.(\ref{eq-9}) on the clustering performance.
In our method, $\alpha$ determines the fusion ratio between the graph autoencoding module and the ZINB-based autoencoding module. 
As shown in the figure, as $\alpha$ increases from 0.1 to 0.5, both NMI and ARI values remain relatively stable, reaching their highest at 0.5. 
This indicates that our method is robust to changes in $\alpha$ within this range. However, when $\alpha$ increases further, the performance drops significantly, demonstrating the effectiveness of our fused embedding strategy. 
Therefore, $\alpha$ is set to 0.5 in our experiments.
\begin{figure*}[htp]%
    \centering
    \includegraphics[width=1\textwidth]{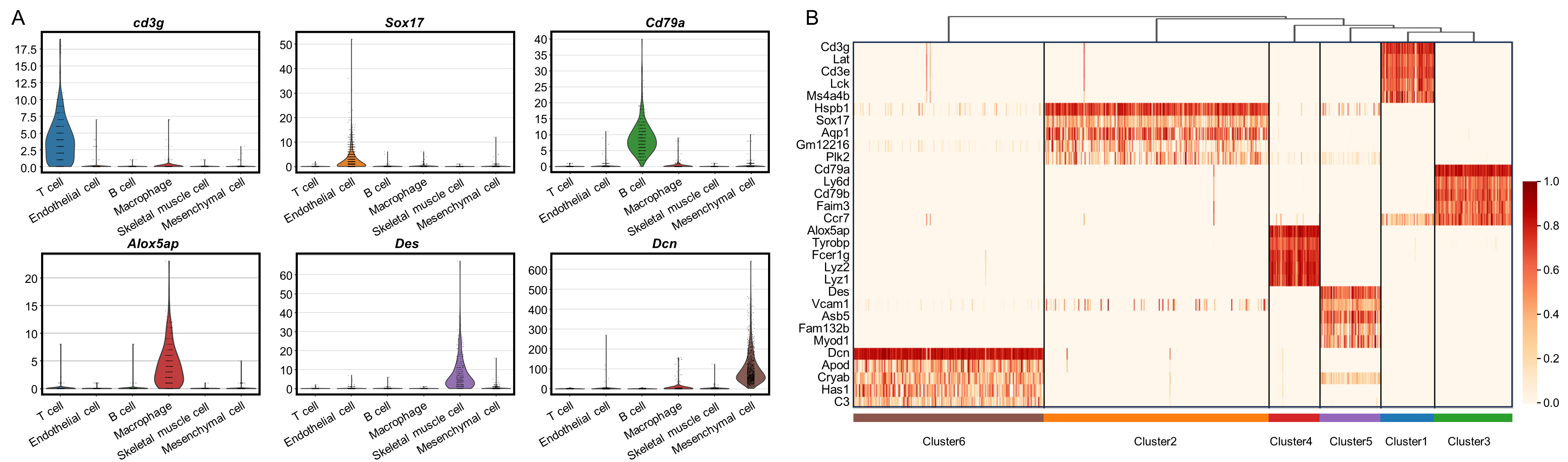}
    \caption{(A) depicting the expression distribution of six representative genes significantly expressed in different cell types. Each violin plot illustrates the expression levels of a specific gene across various cell types, highlighting the distinct expression of these genes in individual cell clusters.
    (B) Heatmap displaying the expression patterns of multiple genes across different cell types. Each row in the heatmap represents a gene, while each column represents a single cell, revealing the expression pattern of specific genes within each cell cluster.}\label{fig-4}
\end{figure*}
\subsubsection{Influence of Highly Variable Genes}
The number of highly variable genes is a critical parameter influencing the richness of data representation. 
\autoref{fig-2}B illustrates the average clustering performance of scASDC across six datasets with varying numbers of highly variable genes (500, 1000, 1500, and 2000).
The results indicate that both NMI and ARI values improve as the number of highly variable genes increases, with the best performance observed at 2000 genes. 
This suggests that using a larger number of highly variable genes offers a more comprehensive representation of the data, resulting in better clustering outcomes.
\subsubsection{Influence of KNN Parameter $k$}
\autoref{fig-2}C shows the effect of varying the KNN parameter $k$ on the clustering performance across six datasets. The NMI and ARI values are plotted against different $k$ values ranging from 5 to 25. The results show that performance peaks at $k = 10$ and $k = 20$, with a noticeable dip observed at $k = 15$. Consequently, we set $k$ to 10 in our experiments.

\subsection{Ablation studies}
To evaluate the effectiveness of different components within the scASDC framework, we conduct an ablation study. The study focuses on the impact of removing three key elements: ZINB loss, the attention mechanism, and graph loss.

Specifically, in the ZINB-based autoencoder module, we remove the ZINB loss $\mathcal{L}_{ZINB}$, referred to as scASDC w/o ZINB Loss. In the layer-by-layer embedding of the encoder, we remove the attention mechanism and sum the representations of the two autoencoder modules, referred to as scASDC w/o Attention. We also remove the graph loss $\mathcal{L}_{GAE_{A}}$, referred to as scASDC w/o Graph Loss.

We repeat each method five times and take the average. As shown in \autoref{tab-3}, the removal of the ZINB module, the attention mechanism, and the graph structure all negatively impact the clustering performance of the model. This demonstrates that each module in our scASDC model is both reasonable and effective.

\subsection{Downstream analysis}
We visualize the clustering results of the algorithms on the six datasets using UMAP for two-dimensional visualization, as shown in \autoref{fig-3}. The figure demonstrates that our scASDC method more effectively separates cells of different populations compared to other algorithms.

To verify whether the embedding representation obtained by the scASDC method can facilitate functional genomics interpretation, we perform a series of functional genomics analyses on the Qx Limb Muscle dataset. The results show that the embedding representation obtained by our method retains important expression patterns and structural information from the original gene expression data.
\begin{figure}[!htp]%
    \centering
    \includegraphics[width=0.49\textwidth]{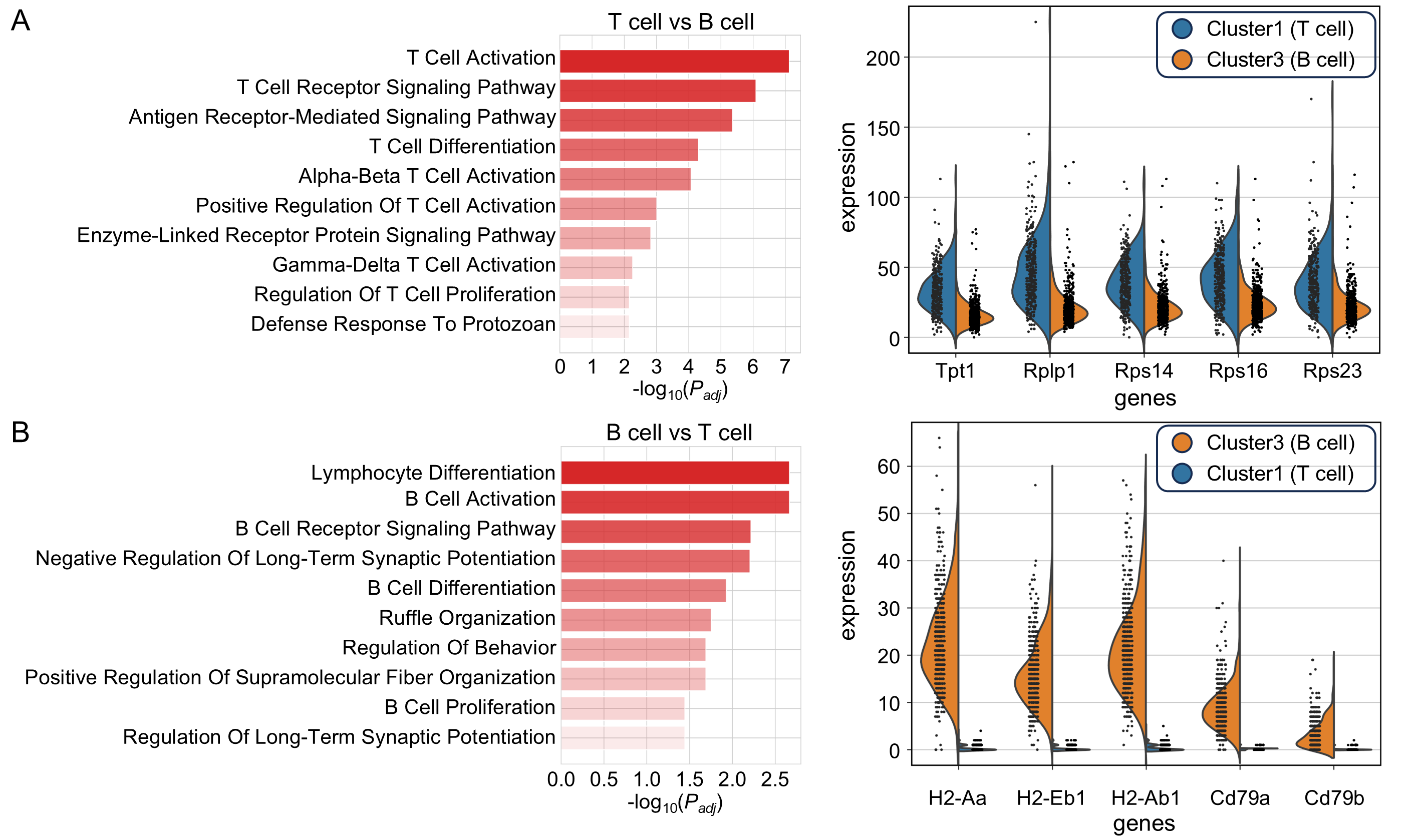}
    \caption{Gene function pathway analysis and expression visualization of Cluster1 (T cells) vs Cluster3 (B cells) identitied by our scASDC.}\label{fig-5}
\end{figure}

Specifically, we use the scanpy package to identify the most significantly differentially expressed genes in each cluster compared to other clusters and illustrate the detailed distribution of these genes across different clusters, as shown in \autoref{fig-4}A. The results indicate that our scASDC method identifies highly expressed genes in each cluster and retains this valuable information in the embedding representation. Furthermore, we plot the expression heat map of the top five differentially expressed genes in each cell cluster across all cells (\autoref{fig-4}B). From the heat map, we can clearly observe the expression differences of these genes in different cell clusters.

Furthermore, we perform functional enrichment analysis on the two cell clusters (T Cell and B Cell) of the Qx Limb Muscle dataset. In \autoref{fig-5}A, the enrichment analysis reveals that the T cell-related cluster (Cluster1) is significantly enriched in immune-related biological processes, including T cell activation, T cell receptor signaling pathways, and antigen receptor-mediated signaling pathways. Conversely, \autoref{fig-5}B illustrates that the B cell-related cluster (Cluster3) is significantly enriched in processes such as B cell activation, B cell receptor signaling pathways, and lymphocyte differentiation. The violin plots on the right side of the figure depict the expression distribution of representative genes specifically expressed in the two clusters, further corroborating the functional distinctions between these clusters.

\section{CONCLUSIONS}\label{CONCLUSIONS}
In this study, we propose a deep learning method, scASDC, for the effective clustering of scRNA-seq data by integrating content and structural information. Our approach leverages a ZINB-based autoencoder and a graph autoencoder, fused through an attention mechanism, to create a robust representation of the data. This method effectively captures both gene expression patterns and the structural relationships between cells. Through extensive experiments on six scRNA-seq datasets, scASDC demonstrates superior performance compared to seven baseline methods, achieving the highest average clustering indices. Our ablation study highlights the significance of each component in scASDC, with the complete model outperforming variations lacking ZINB loss, the attention mechanism, or graph loss. Visualization of clustering results using UMAP further confirms the capability of scASDC to distinctly separate different cell populations. Moreover, pathway enrichment analysis reveals meaningful biological insights, with significant pathways identified for both T cells and B cells. This underscores the ability of scASDC to not only cluster cells accurately but also facilitate downstream biological interpretation.

\section*{Acknowledgment}
The work was supported in part by the National Natural Science Foundation of China (62262069), in part by the Yunnan Fundamental Research Projects under Grant (202201AT070469) and the Yunnan Talent Development Program - Youth Talent Project.
\balance
\small
\bibliography{references.bib} 
\bibliographystyle{IEEEtran}  
\end{document}